\documentclass[10pt, conf, twocolumn]{IEEEtran}
\usepackage{graphicx} % for pdf, bitmapped graphics files
\usepackage{amsmath} % assumes amsmath package installed
\usepackage{amssymb}  % assumes amsmath package installed

\newtheorem{theorem}{\textbf{Theorem}}

\newtheorem{proof}{\textbf{Proof}}
\newtheorem{corollary}{\textbf{Corollary}}
\usepackage{caption}
\usepackage{subcaption}
\usepackage{color}
\usepackage{algorithm}
\usepackage[noend]{algpseudocode}

\makeatletter
\def\BState{\State\hskip-\ALG@thistlm}
\makeatother

\title{\LARGE \bf
Adaptive and Resilient Revenue Maximizing Dynamic Resource Allocation and Pricing for Cloud-Enabled IoT Systems
}

\author{Muhammad Junaid Farooq and Quanyan Zhu \vspace{-0.2cm}% <-this % stops a space
\thanks{The authors are with the Department of Electrical \& Computer Engineering at the Tandon School of Engineering, New York University (NYU), Brooklyn, NY 11201, USA. Emails: {\{\tt\small mjf514,qz494\}@nyu.edu}.
\indent This research is partially supported by a DHS grant through Critical Infrastructure Resilience Institute (CIRI), grants CNS-1544782 and SES-1541164 from National Science of Foundation (NSF).
}%
}

\begin{document}

\maketitle
\thispagestyle{empty}
\pagestyle{empty}

%%%%%%%%%%%%%%%%%%%%%%%%%%%%%%%%%%%%%%%%%%%%%%%%%%%%%%%%%%%%%%%%%%%%%%%%%%%%%%%%
\begin{abstract}

Cloud computing is becoming an essential component in the emerging Internet of Things (IoT) paradigm. The available resources at the cloud such as computing nodes, storage, databases, etc. are often packaged in the form of virtual machines (VMs) to be used by remotely located IoT client applications for computational tasks. However, the cloud has a limited number of VMs available and hence, for massive IoT systems, the available resources must be efficiently utilized to increase productivity and subsequently maximize revenue of the cloud service provider (CSP). IoT client applications generate requests with computational tasks at random times with random complexity to be processed by the cloud. The CSP has to decide whether to allocate a VM to a task at hand or to wait for a higher complexity task in the future. We propose a threshold-based mechanism to optimally decide the allocation and pricing of VMs to sequentially arriving requests in order to maximize the revenue of the CSP over a finite time horizon. Moreover, we develop an adaptive and resilient framework that can counter the effect of realtime changes in the number of available VMs at the cloud server, the frequency and nature of arriving tasks on the revenue of the CSP.

\end{abstract}

%%%%%%%%%%%%%%%%%%%%%%%%%%%%%%%%%%%%%%%%%%%%%%%%%%%%%%%%%%%%%%%%%%%%%%%%%%%%%%%%
\section{INTRODUCTION}

In recent years, due to the ubiquity of the internet, there has been an increasing trend towards offloading computing, control, and storage to the cloud instead of doing it locally at the client side~\cite{zhu}. This trend is expected to accentuate with the proliferation of the Internet of things (IoT)~\cite{iot,twc}. The IoT applications can request for cloud resources for a variety of different computational tasks. For instance, they can invoke machine learning and data analytics models already implemented in the cloud server to enable powerful features such as predictive analytics, video processing, and natural language processing. With a massive surge in the number of applications requesting the cloud for computational resources in the future, there will be a need for an efficient allocation and pricing mechanism at the cloud server. A cloud service provider (CSP) has several resources such as computing nodes, storage, databases, etc., that can be used remotely by IoT applications. Often, these resources are packaged into virtual machine (VM) instances that act as processing units. When the number of requesting applications is large and the available VMs are limited, as envisioned in massive IoT systems~\cite{massive_iot}, it is important to select which applications are serviced particularly when the allocation is planned for longer periods of time.

\begin{figure}[t]
  \centering
  \includegraphics[width=3.0in]{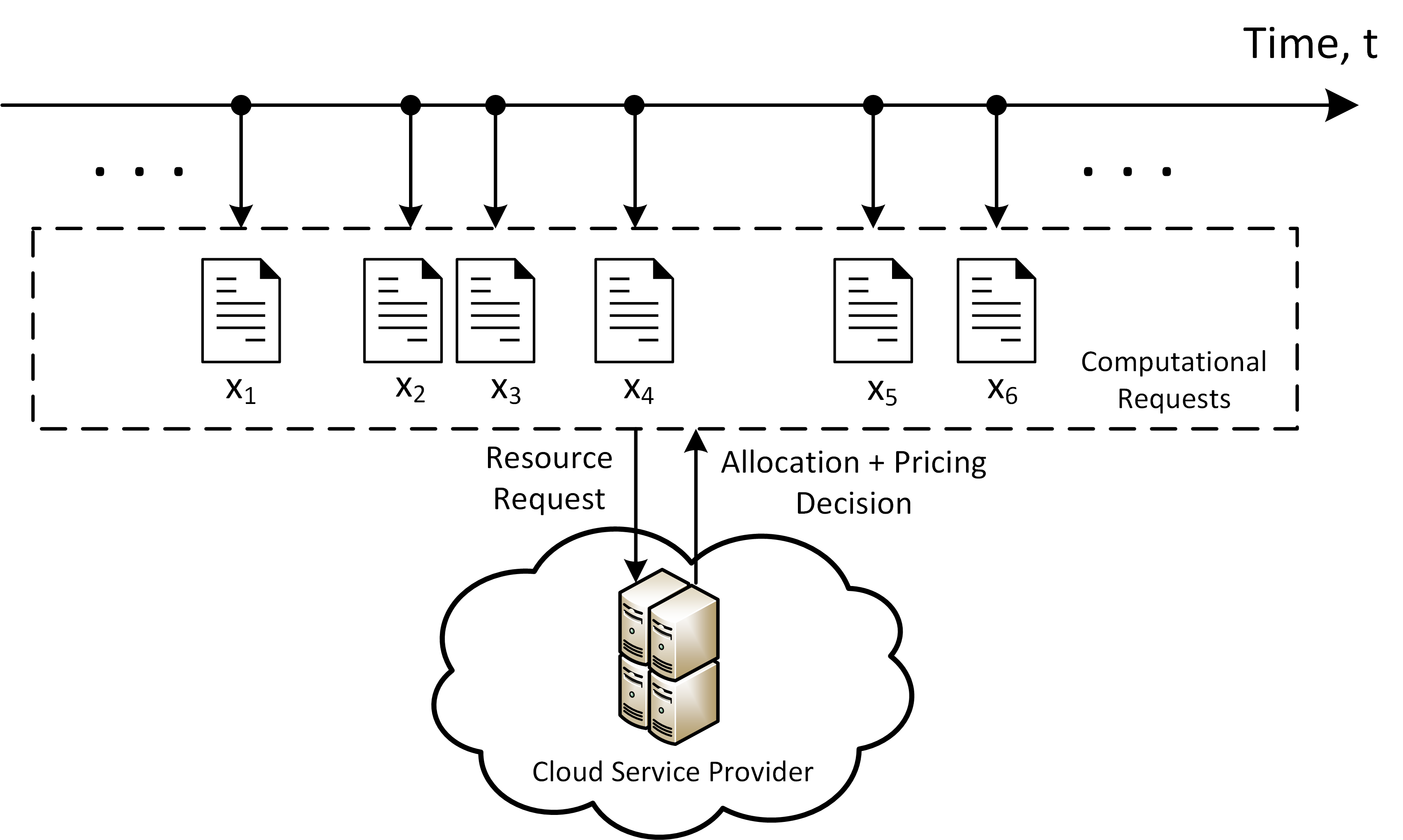}
  \caption{Cloud service provider allocating available VMs to sequentially arriving computational requests by IoT client applications.\vspace{-0.3in}}\label{fig:sys_model}
\end{figure}

The challenges faced by the CSP in allocating the available VMs to requesting applications are twofold. Firstly, the available VMs at the CSP are limited, so it is important to allocate the most computationally intensive tasks to the available VMs in order to maximize the productivity of the IoT client applications and the generated revenue by charging them appropriately. However, the tasks arrive sequentially at the server and the CSP has to decide immediately to allocate a VM to it or to wait for a more valuable task in the immediate future. The challenge lies in the uncertainty about the nature of upcoming requests in the future. A computationally intensive task may not ever request for service while the low complexity tasks are refused service. It leads to an under-utilization of resources resulting in lower productivity and revenue of the CSP. On the other hand, if the VMs are allocated to low complexity tasks, then a high complexity task may request service in the future and has to be refused due to the unavailability of an VM at the cloud server.
%Secondly, as the time progresses, if the available VMs are not allocated while waiting for computationally intensive applications, the CSP will also lose the opportunity of generating revenues from low complexity tasks by allocating them to the available VMs.
Therefore, there is a need for a dynamically efficient mechanism for allocating and pricing the VMs that takes these tradeoffs into account. Fig.~\ref{fig:sys_model} illustrates the sequential arrival of IoT computation requests at the CSP.

%\subsection{Related Work}
There has been considerable work in the literature towards resource allocation in cloud computing environments~\cite{resource_allocation_cloud}. The focus is mainly on efficient resource management and load balancing for higher availability and performance~\cite{cloud_performance} or resource allocation and pricing for revenue maximization~\cite{risk_based,cloud_pricing}. Regarding dynamic pricing and revenue maximization, several works exist such as~\cite{spot_markets} which use price control to adjust demand levels. Others have used auction mechanisms to collect bids and allocate available computational resources such as~\cite{cloud_auction}. However, most existing works on resource allocation in cloud computing do not take into account the sequential arrival of computing tasks and the uncertainty about the future. This is essential in the setting of cloud computing because the computational requests are spontaneous and the decision for allocation has to be made immediately upon arrival. A dynamically efficient policy for allocating resources to sequentially arriving agents in order to maximize social welfare was first proposed by Albright~\cite{albright}. Consequently, a revenue maximizing approach towards sequential allocation of resources has been introduced in~\cite{revenue_maximization}. However, their work deals with heterogeneous resources and cannot be used to model situations with identical resources. In cloud-enabled IoT systems, often multiple identical resources such as VMs are available to be allocated to client applications. A framework for pricing the cloud for maximizing revenue is proposed in~\cite{cloud_pricing}. However, their solution is based on stochastic dynamic programming which cannot adapt in realtime scenarios. Our solution provides a dynamically optimal plug and play policy that can be pre-computed and used in realtime using a lookup table.

%\subsection{Contribiutions}
In this paper, we develop an adaptive and resilient dynamic resource allocation and pricing framework for cloud-enabled IoT systems. We present an optimal dynamic policy to filter incoming service requests by IoT applications based on the complexity of the tasks.
%It does not depend on the characteristics of the available VMs. It only depends on the number of available VMs and the statistical properties of the computational tasks at the cloud.
The qualification threshold for tasks is adaptive to the number of available VMs, the arrival rate of requests, and their average complexity. The optimal policy can be dynamically updated in order to maintain high expected revenues of the CSP.
%For instance, if the computational requests are less frequent, then the proposed framework adapts by reducing the qualification threshold and hence the price so that a higher proportion of the incoming requests can be serviced to generate maximum revenue. On the other hand, if the frequency of requests is very high, then it adapts by increasing the qualification threshold and price. Consequently, the CSP will wait for high complexity tasks to arrive and make higher revenues.
Furthermore, the proposed framework is also able to adapt according to the changing availability of the VMs due to reprovisioning of resources for other applications or due to the effect of malicious attacks.

%The VMs may become unavailable due to other requirements at the cloud. Similarly, additional VMs can be created and provisioned for allocation to computational requests.
%The rest of the paper is organized as follows: Section~\ref{Sec:Syst_model} describes the system model, Section~\ref{Sec:Methodology} presents the methodology used to derive revenue maximizing policies, and Section~\ref{Sec:Resilience} explains the proposed adaptive and resilient allocation and pricing framework. The numerical results and discussion are provided in Section~\ref{Sec:Results}. Finally Section~\ref{Sec:Conclusion} concludes the paper with some future directions.

\section{System Model}\label{Sec:Syst_model}
We consider a CSP having a set of $N_t \in \mathbb{Z}^+$ available VMs at time $t\in[0,T]$. The VMs are identical and are characterized by their computational efficiency\footnote{The computational efficiency can be determined by evaluating the relative time taken by the VM to successfully execute a benchmark task.} denoted by $q \in [0,1]$. The CSP receives requests\footnote{Throughout the paper, we use the word `requests' to refer to computational tasks generated by IoT applications that arrive at the CSP for processing.} for computation by IoT client applications. These requests arrive sequentially at the cloud server according to a Poisson process with density $\lambda \in \mathbb{R}^+$ requests per unit time. Each task has computational complexity denoted by $X \in \mathbb{R}^+$. The computational complexity can be measured in terms of the number of CPU cycles or equivalently the time required to complete a given computational task. The computational complexities of sequentially arriving tasks are considered to be independent and identically distributed (i.i.d.) random variables with probability density function (pdf) and cumulative distribution function (cdf) denoted by $f_X(x)$  and $F_X(x)$ respectivley.

The utility of the $i^{\text{th}}$ arriving client application with a task complexity of $x_i, i \in \mathbb{Z}^+,$ that is allocated a VM with efficiency $q$, is measured by the product $qx_i$, which refers to the resulting value created by the allocation or the productivity.
%The total expected productivity of the clients application if $N_t$ VMs were available for allocation at time $t$ can be expressed as follows:
%\begin{align}
%U(\{q\}_{N_{t}},t) = \mathbb{E} \sum_{i=1}^{N_t} q x_j,
%\end{align}
%where $x_j$ is the task allocated to the $i^{\text{th}}$ available VM.
Since the available VMs are limited, the CSP needs to allocate the VMs to only the high complexity arriving tasks in order to increase the total productivity of the clients as well as efficient utilization of available computational resources. Creating higher value sets the ground for the CSP to charge higher prices and hence generate more revenue.
However, the decision has to be taken immediately\footnote{We assume that the tasks are impatient and need to be processed immediately without delay.} upon arrival of the tasks without knowledge of tasks arriving in the future. Therefore, the CSP has the option to either allocate one of the available VMs or to refuse the requesting application.

\subsection{Allocation \& Pricing Rule}
In order to allocate available VMs to randomly arriving computational requests, we adapt the result from the sequential stochastic assignment literature, which is based on the Hardy-Littlewood-Polya inequality~\cite{hard_book} and is stated by the following theorem:
\begin{theorem}[Adapted from~\cite{albright}]
{\em If there are $n$ VMs with computational efficiencies $q_1, q_2, \ldots, q_n$ such that $0 < q_1 \leq q_2 \leq \ldots \leq q_n$, then there exists a set of functions
\begin{align*}
0 = z_{n+1}(t) \leq z_{n}(t) \leq \ldots \leq z_1(t) \leq z_0(t) = \infty.
\end{align*}
such that it is optimal (in terms of social welfare) to assign a VM with efficiency $q_i$ to an incoming task with complexity $x$ if $z_{n-i+1}(t) \leq x \leq z_{n-i}(t)$. Furthermore, if $x < z_{n}(t)$, it is optimal not to allocate it.
}
\end{theorem}

In the case of identical objects, the CSP needs to set only a single dynamic threshold, which we refer to
as the \emph{qualification threshold}, that allows it to decide whether to allocate a VM or not based on the nature of the arriving task. Let $y_{N_t}(t) \in \mathbb{R}^+, \forall t \in [0,T]$ denote the threshold if $N_t$ VMs are available for allocation at time $t$.
In other words, only the requests with $x_i \geq y_{N_t}(t), i \in \mathbb{Z}^+$, will be allocated to an available VM at time $t$. %We assume long term allocation, i.e., once a VM is allocated to a computational task, it cannot be allocated to another task in the future.
The allocation process has to be completed within a finite time horizon denoted by $T$. Since the VMs have no commercial value if they remain idle or unallocated during the allocation period, therefore the threshold needs to be dynamic in order to efficiently generate revenue from available resources. The decision problem lies in that fact that it may be more valuable to assign a VM to a low complexity task than waiting for a high complexity task to arrive in the future which may not ever realize.

%The CSP sets a qualification threshold denoted by $y_{N_t}(t)$ on the minimum type of task to be allocated a VM.

%This makes the decision problem challenging as the CSP needs to decide without knowing the type of tasks that will arrive in the future.
The next step is to develop a pricing scheme for the available VMs. Since all the VMs are identical in terms of their performance, therefore they must be priced equally. The threshold based allocation policy provides a natural method for pricing the available VMs. Since each arriving task that is successfully allocated a VM at time $t$ receives a value of at least $qy_{N_t}(t)$. Therefore, it is fair to charge the price $\mathcal{P}: [0,1] \times \mathbb{R}^+ \times [0,T]\rightarrow \mathbb{R}^+$ to a qualified task for being processed by a VM as follows: \vspace{-0.1cm}
\begin{align}
\mathcal{P}(q,y_{N_t}(t),t) = q y_{N_t}(t) + S(t),
\end{align}
where $S(t)$ represents the constant additional pricing independent of the allocation. This pricing policy is implementable as any individually rational client will be willing to pay at least an amount equal to its received value. Note that $S(t)$ can be used to adjust the prices due to external factors such as promotions, packages, pricing agreements, etc. In the following section, we provide the dynamically optimal allocation threshold and the resulting price charged by the CSP to allocated tasks.
%arriving requests will be willing to pay the value that they obtain. Since any qualifying request that is served by the CSP obtains a utility of at least $qy_{N_0}(t)$, so it satisfies the individual rationality constraints.

%Since all VMs are identical, there should be a single threshold that serves as a decision rule of whether to allocated a VM to an upcoming request or not. This also ensures that any two qualifying requests that arrive at the same time obtain the same computational resource and pays the same price.

\section{Methodology} \label{Sec:Methodology}
We will begin by quantifying the total expected revenue of the CSP and subsequently derive the optimal dynamic threshold that maximizes the revenue. The total expected revenue of the CSP $\mathcal{R}: [0,1] \times \mathbb{R}^+ \times [0,T]\rightarrow \mathbb{R}^+$ if $N_t$ identical VMs with computational efficiency $q$ are available at time $t$ and a qualification threshold $y_{N_t}(t)$ is used from time $t$ onwards can be expressed as follows: \vspace{-0.2cm}
\begin{align} \label{revenue_single_new}
\mathcal{R}(q, y_{N_t}(t),t) = \sum_{n=1}^{N_t} \ q \int_t^T y_{N_t}(s) h_{n}(s) ds + \kappa_T, \vspace{-0.2cm}
\end{align}
where $h_{n}(t)$ is the density of waiting time until the $n^{\text{th}}$ arrival of a qualifying task, i.e., having a task complexity greater than $y_{N_t}(t)$, and $\kappa_T$ is a constant factor due to the additional pricing function $S(t)$. The density of waiting time can be expressed by the density of the $n^{\text{th}}$ arrival in a non-homogeneous Poisson process with intensity $\hat{\lambda}(s) = \lambda (1 - F(y_{N_t}(s)))$. Consequently, the density can be written as follows~\cite{stochastic_book}:\vspace{-0.6cm}

{\small
\begin{align}
h_{n}(s) &= \hat{\lambda}(s) \exp \left(- \int_t^s \hat{\lambda}(u) du \right) \frac{\left(\int_t^s \hat{\lambda}(u) du \right)^{n-1}}{(n-1)!} , t \leq s \leq T.
\end{align}
}
The objective is to select a time-varying threshold $y_{N_t}(t)$ which maximizes the expected revenue functional given by~\eqref{revenue_single_new}. The problem can be formally stated as follows:
\begin{equation*}
\begin{aligned}
& \underset{y_{N_t}(t)}{\text{maximize}}
& & \mathcal{R}(q, y_{N_t}(t),t) \\
& \text{subject to}
& & y_{N_t}(t) \geq 0, \forall t\in [0,T].
\end{aligned}
\end{equation*}
Note that the optimization is over the space of functions where an optimal function $y_{N_t}(t)$ is sought for a given number of available VMs at time $t$.
%The problem in choosing the qualification threshold is as follows. If the threshold is set too low, then there is a higher probability of requests qualifying but the generated revenue will be small. On the other hand, if the threshold is set too high, then the probability of a request qualifying will be low but the earned revenue will be high.
Our aim is to design the threshold function that strikes the optimal balance between the number of qualifying tasks and the generated revenue. In the sequel, we provide the optimal qualification threshold for maximum revenue generation by the CSP and the properties of the optimal policy.
%Moreover, based on the optimal policy, we propose an adaptive and resilient allocation and pricing framework that can react to the changes in the number of available VMs at the cloud server.

%In this section, we provide aframework for dynamically maximizing the average expected QoE of the IoT ecosystem and the revenue of the CSP. We focus on a mechanism design approach, which is based on developing optimal implementable policies for allocation and pricing. In the subsequent subsections, a direct mechanism is provided whereby each requesting application reports its minimum required response rate and the CSP allocates one of the available cloud nodes to it, i.e., forwards the received data to either the main cloud server or one of the available fog nodes. For the optimal allocation to be implementable, an allocation policy and a payment rule is required that is incentive compatible\footnote{Incentive compatibility is a concept from mechanism design theory that ensures that no agent has an incentive to misreport its privately known characteristic.}~\cite{Myerson} in the presence of individually rational users. We first state an allocation rule which satisfies the aforementioned conditions and then provide a pricing strategy that subsequently implements the allocation.

\begin{theorem} \label{main_theorem}
{\em If $N_t$ VMs are available to the CSP at time $t$, computational requests arrive sequentially according to a Poisson process with intensity $\lambda$ and the computational complexity of tasks are i.i.d. random variables with pdf $f_X(x)$ and cdf $F_X(x)$, then it is optimal to allocate an available VM to an incoming computational request if the complexity of an upcoming task $x \geq y_{N_t}(t)$. The optimal $y_{N_t}(t)$ satisfies the following integral equation:\vspace{-0.4cm}

{\small
\begin{align} \label{eq_main_theorem}
y_{N_t}(t) \hspace{-0.05cm} =& \frac{1 - F_X(y_{N_t}(t))}{f_X(y_{N_t}(t))} \hspace{-0.05cm} +  \lambda \hspace{-0.05cm} \int_t^T \hspace{-0.1cm}\frac{(1 - F_X(y_{N_t}(s)))^2 }{f_X(y_{N_t}(s))} J_{N_t}(t,s)ds,
\end{align}
}
where $J_{N_t}(t,s)$ can be expressed as follows:
\small
\begin{align}
&J_{N_t}(t,s) = \frac{1}{\sum_{n=1}^{N_t} \frac{1}{(n-1)!}  \left(\int_t^s \hat{\lambda}(u) du\right)^{n-1} } \times \notag\\ &\sum_{n=1}^{N_t} \frac{1}{(n-1)!} \left( \left(\int_t^s \hat{\lambda}(u) du\right)^{n-1} \hspace{-0.3cm} - (n-1)\left(\int_t^s \hat{\lambda}(u) du\right)^{n-2}\right).
\end{align}
}
\begin{proof}
See \textbf{Appendix~\ref{proof_main_theorem}}.
\end{proof}
\end{theorem}

%\begin{theorem}
%The optimal revenue maximizing threshold exists if $y_i(s) - \frac{1 - F_X(y_i(s))}{f_X(y_i(s))}$ is increasing in $y_i(s)$.
%\begin{proof}
%Contraction Mapping
%\end{proof}
%\end{theorem}

%Note that the optimal qualification threshold depends only on the number of available VMs and is independent of their efficiency. However, the efficiency is important for complete characterization of the expected revenue. The qualification threshold also depends on the statistical properties of the sequentially arriving tasks.
The behaviour of the optimal dynamic threshold for large number of available VMs is provided by the following corollary.
\begin{corollary}
{\em If the number of available VMs is large, then the revenue maximizing threshold becomes constant and the allocation mechanism reduces to a first price auction mechanism, i.e., allocate a VM to a task if the complexity is higher than the virtual valuation.}\\
\begin{proof}
In the optimal allocation policy, if we let $N_t \rightarrow \infty$, then the optimal threshold solves the following integral equation:\vspace{-0.4cm}

{\small
\begin{align}
y_\infty(t) =& \frac{1 - F_X(y_{\infty}(t))}{f_X(y_{\infty}(t))} + \int_t^T \frac{(1 - F_X(y_{\infty}(t)))^2}{f_X(y_{\infty}(t))} \times \notag \\ & \qquad \qquad \qquad \qquad \qquad  \left( \underset{N_t \rightarrow \infty}{\lim} J_{N_t}(t,s) \right) ds.
\end{align}
}
Now, $\underset{N_t \rightarrow \infty}{\lim} J_{N_t}(t,s)$ can be evaluated as follows:\vspace{-0.4cm}

{\small
\begin{align}
\hspace{-0.1cm}\underset{N_t \rightarrow \infty}{\lim} J_{N_t}(t,s) &= \frac{   \sum_{n=1}^{\infty} \frac{H^{n-1}(s)}{(n-1)!}   -  \sum_{n=1}^{\infty} \frac{(n-1)H^{n-2}(s)}{(n-1)!}}{\sum_{n=1}^{\infty} \frac{H^{n-1}(s)}{(n-1)!}} \notag \\&= \frac{e^{H(s)} - e^{H(s)}  }{e^{H(s)}} = 0.
\end{align}
}
Therefore, it follows that $y_\infty(t) = \frac{1 - F_X(y_{\infty}(t))}{f_X(y_{\infty}(t))}$. Note that $x - \frac{1 - F_X(x)}{f_X(x)}$ is referred to as the virtual valuation of the agent of type $x$ in mechanism design literature~\cite{Myerson:1981:OAD:2781650.2781656}. Hence, it can be concluded that if the number of available VMs is large, then only the virtual valuation of the arriving tasks can be recovered and the CSP is willing to offer the VMs for lowest possible threshold.
\end{proof}
\end{corollary}

%The existence of a unique solution to the integral equation in Theorem~\ref{main_theorem} is stated in the following lemma.
%\begin{lemma}
%{\em There exists a unique optimal threshold function $y_{N_t}(t), t \in \mathbb{R}^+, N_t \in \mathbb{Z}^+$ exists that satisfies the integral equation~\eqref{eq_main_theorem}.\\
%\begin{proof}
%See Appendix
%\end{proof}
%}
%\end{lemma}

The behaviour of the dynamically optimal qualification threshold with a variation in the number of available VMs at time $t$ can be summarized by the following theorem.
\begin{theorem}\label{theorem2}
{\em The qualification threshold of the tasks and consequently price of VMs decreases as the number of available VMs at the cloud server increases and vice versa, i.e., $y_{M_t}(t) \leq y_{N_t}(t)$ if $M_t \geq N_t, \forall t$.}\\
\begin{proof}
See \textbf{Appendix~\ref{proof_theorem2}}.
%Comparison lemma. The solution to the differential inequality is bounded above by the solution of the equality.
%The allocation thresholds are decreasing with the number of available VMs, i.e., $y_m(t) \leq y_n(t), \forall t$ if $m \leq n$.
%Since threshold decrease, so from equation () price also decreases.
\end{proof}
\end{theorem}

In the following section, we discuss how the dynamically optimal policy leads to an adaptive and resilient behaviour in the revenue of the CSP and describe the developed mechanism algorithmically.

\section{Adaptive and Resilient Allocation and Pricing Policy}\label{Sec:Resilience}
The number of available VMs at the cloud server may change over time as some of them might become unavailable due to failure or malicious attacks~\cite{cloud_attack}. The CSP might also destroy the created VMs in order to free up computational resources for other applications. On the other hand, previously allocated VMs might be released by applications or new VMs may be provisioned by the CSP in real time to accommodate higher demand. However, a change in the available number of VMs may affect the expected revenue of the CSP under a particular allocation and pricing policy particularly if there is a significant decrease in the number of remaining VMs. In order to reduce any negative impact on the expected revenue of the CSP, the proposed revenue maximizing framework will react to the changes in the available number of VMs by adapting the qualification threshold or equivalently the price.

Furthermore, the developed framework can react to changes in the frequency and the nature of computational requests. The optimal resilient policy is denoted by $\Pi(\tilde{\lambda}_t, \tilde{N}_t, t)$, where $\tilde{\lambda}_t$ is the rate of arrival of requests and $\tilde{N}_t$ represents the number of available VMs at time $t$. Note that $\tilde{N}_t = N_t + \eta_t$, where $\eta_t \in \mathbb{Z}$ is the change in the number of available VMs at time $t$. The adaptive qualification threshold $\tilde{y}_{N_t}(t)$ can be pre-computed using the optimal policy framework presented earlier in this Section. The policy then becomes a lookup table that the CSP uses to allocate and price the upcoming tasks. Note that the variations in the inputs of the framework can be directly incorporated into the derived results. For instance, if $\eta_t$ VMs enter/leave the system at time $t$, then it is equivalent to as if there were $N_t + \eta_t$ available VMs at time $t$. Hence, the optimal threshold corresponding to $N_t+\eta_t$ must be used for time $t$ onwards for maximizing revenue.

\begin{figure}[t]
  \centering
  \includegraphics[width=2.5in]{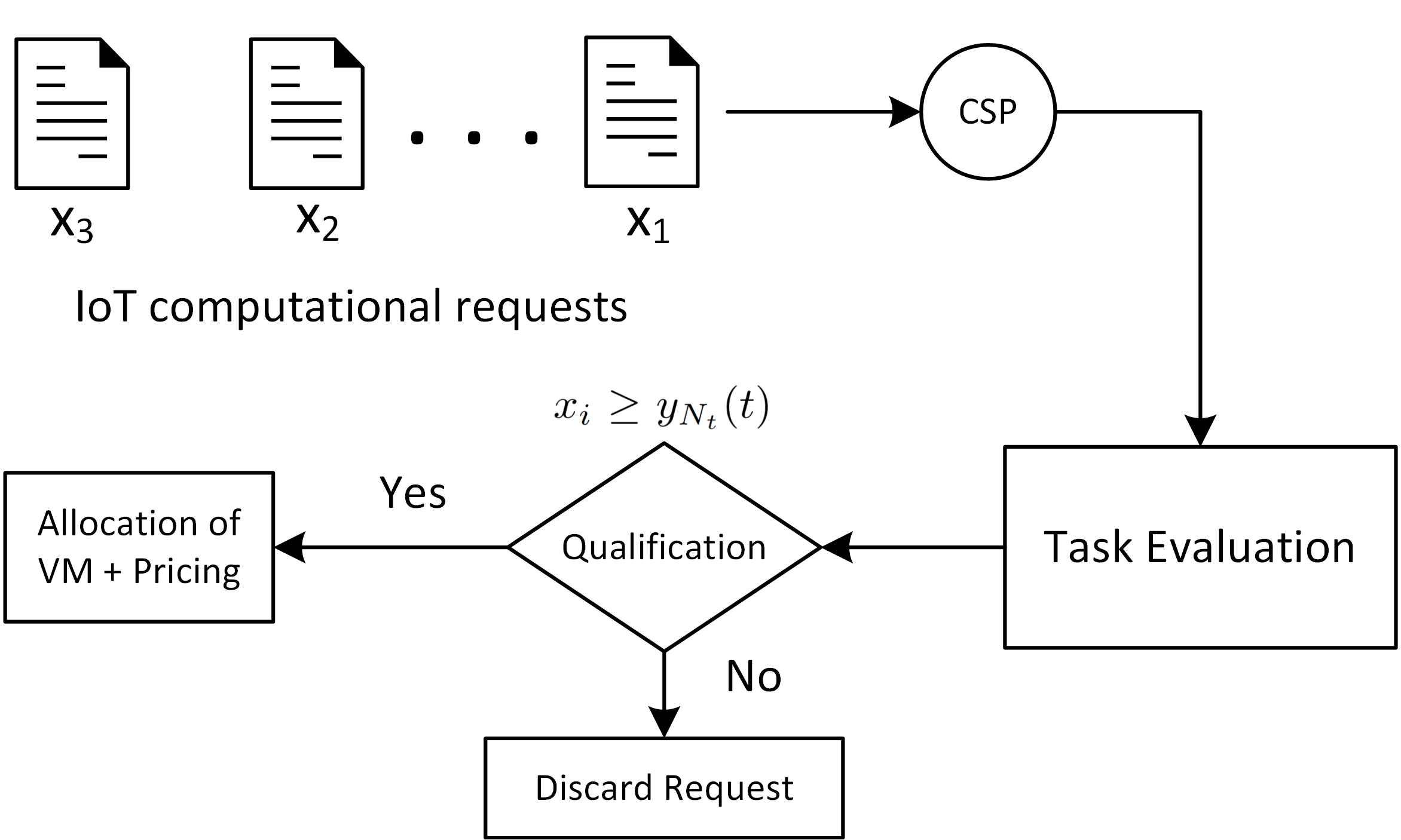}
  \caption{Flow diagram of the adaptive and resilient resource allocation and pricing mechanism.\vspace{-0.15in}}\label{fig:flow_diag}
\end{figure}

%\begin{algorithm}[h!]
%\small
%\caption{Adaptive and Resilient Allocation and Pricing}\label{algorithm1}
%\begin{algorithmic}[1]
%\Require: Task arrival statistics $\lambda$, $f_X(x)$, $F_X(x)$. Initialize $t = 0$.
%\While{$\tilde{N}_t > 0$ \textbf{and } $t \leq T$}
%%\Procedure{}{}
%\If {A computational request arrives}
%\State Evaluate the complexity of the task $\mathrm{x} \gets x_i$.
%\State Determine the number of VMs entering/leaving the system $\eta_t$.
%\State Update the arrival density assumption $\tilde{\lambda}$.
%\State $\tilde{N}_t \gets N_t + \eta_t$.
%\State Obtain qualification criteria using pre-computed lookup table $\Pi(\tilde{\lambda}_t, \tilde{N}_t,t)$, i.e., $\gamma \gets y_{\tilde{N}_t}(t)$.
%\If {$\mathrm{x} > \gamma$}
%\State Allocate one of the VMs to the requesting application.
%\State Charge a price $\mathcal{P}(q,\gamma,t)$.
%\Else
%\State Discard computational request. \Return
%\EndIf
%\Else \
%\Return
%\EndIf
%\EndWhile
%%\EndProcedure
%\end{algorithmic}
%\end{algorithm}

%The adaptive and resilient strategy resource allocation and pricing strategy is summarized in Algorithm~\ref{algorithm1}.
The algorithm proceeds as follows. While the allocation period has not expired and there is still an available VM at the CSP, if an IoT application requests for computation, then the first step is to evaluate the task complexity. Once the complexity is determined, it is compared against the decision threshold. However, the optimal threshold used will depend on the current situation at the CSP. Hence, the updated number of available VMs $\tilde{N}_t$ and the updated arrival rate of requests $\tilde{\lambda}$ is used to read off the optimal policy from the lookup table $\Pi(\tilde{\lambda}_t, \tilde{N}_t, t)$ at time $t$. A flow diagram is provided in Fig.~\ref{fig:flow_diag} to illustrate the sequence of the mechanism.

\section{Numerical Results \& Discussions}\label{Sec:Results}
In this section, we provide numerical results for the proposed adaptive and resilient optimal dynamic allocation and pricing framework. We assume a single CSP having $N_t$ available VMs to allocate to arriving computational requests within an allocation time horizon of $T = 12$ hours. The number of available VMs available at time $t=0$, referred to as $N_0$ is set to be 100. The computational efficiency of the VMs is selected to be $q = 1$ without loss of generality. Note that the characteristics of the VMs is only relevant to the pricing policy but not the allocation.

\begin{figure}
\centering
\begin{subfigure}{.5\linewidth}
  \centering
  \includegraphics[width=1\linewidth]{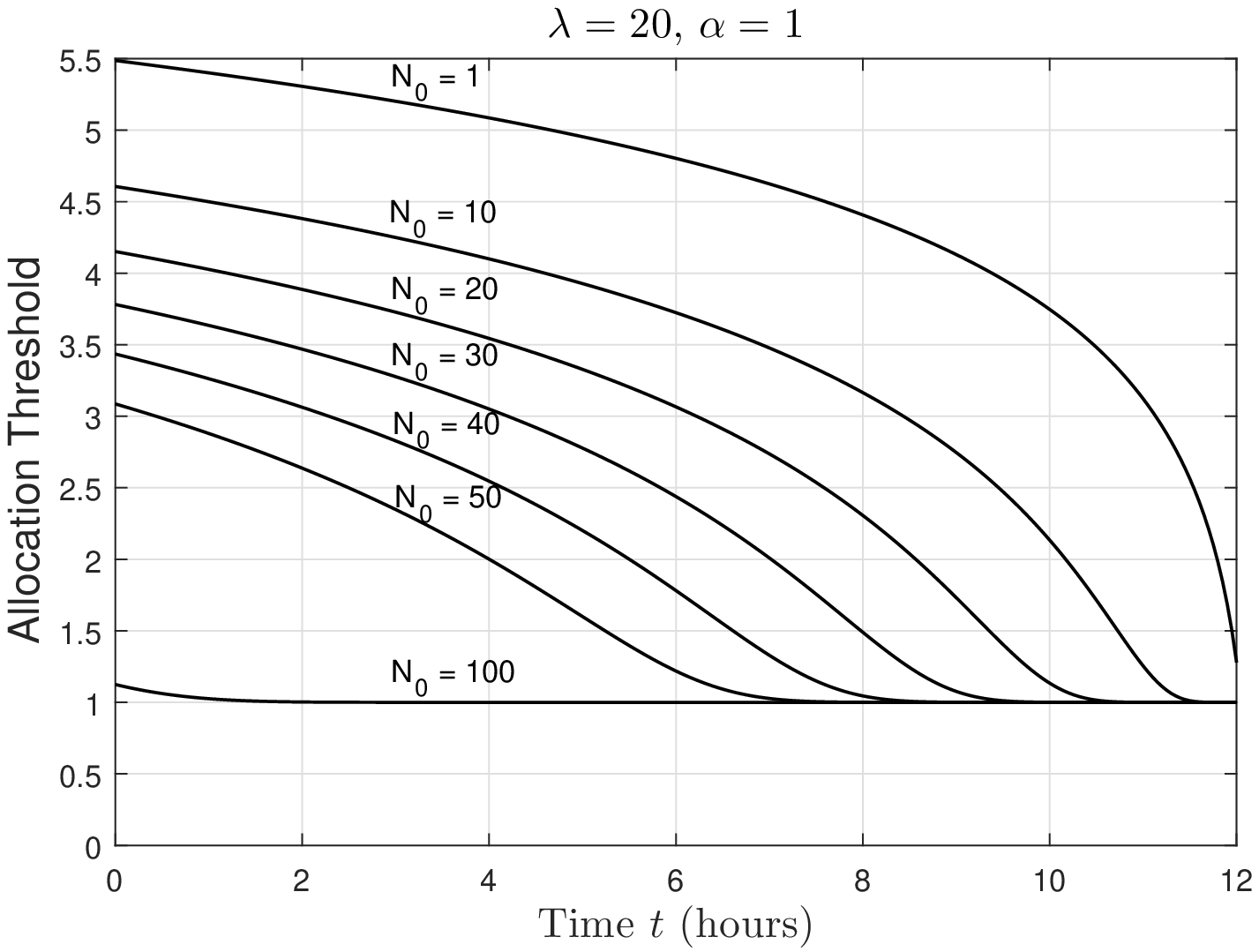}
  \caption{Low frequency of requests.}
  \label{fig:low_rate}
\end{subfigure}%
\begin{subfigure}{0.5\linewidth}
  \centering
  \includegraphics[width=1\linewidth]{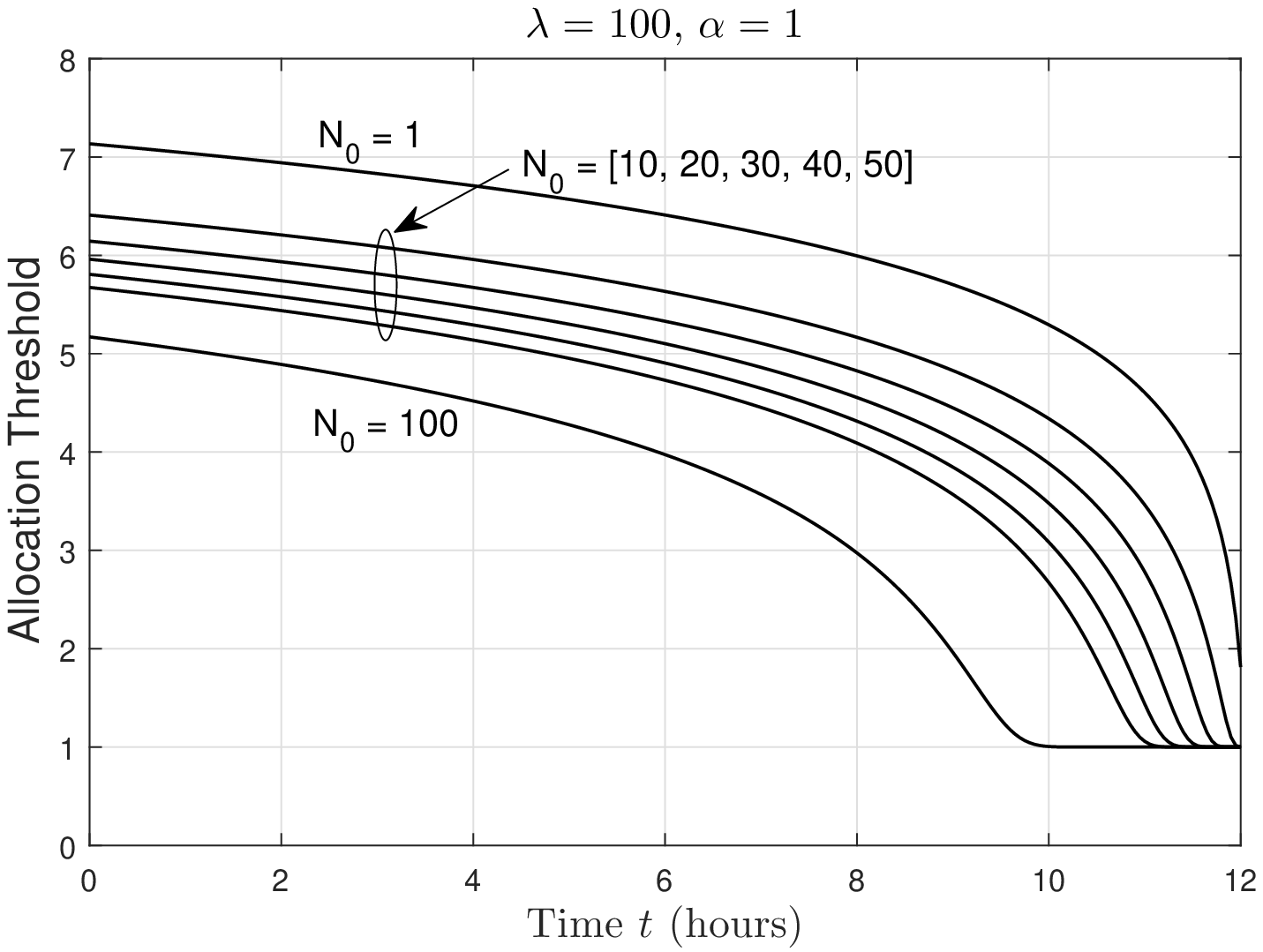}
  \caption{High frequency of requests.}
  \label{fig:high_rate}
\end{subfigure}
\caption{Optimal allocation thresholds for low and high arrival rates of computational requests.\vspace{-0.25in}}
\label{fig:thresholds}
\end{figure}

The tasks arrive at the CSP according to a homogeneous Poisson process with intensity $\lambda = 100$ requests per hour unless otherwise stated.
We also assume that the complexity of sequentially arriving computational requests are distributed according to an exponential distribution with a mean of $\frac{1}{\alpha}$, i.e., $f_X(x) = \alpha e^{-\alpha x}$, and $F_X(x) = 1 - e^{-\alpha x}$. For simplicity, we select $\alpha = 1$, resulting in an average task complexity of 1. The optimal task qualification thresholds in this case if $N_t$ VMs are available at time $t$ can be obtained by the solution of the integral equation expressed by~\eqref{exponential_threshold}. The equation can be solved numerically using the Picard fixed point iteration~\cite{fixed_point}. Fig.~\ref{fig:thresholds} shows the dynamic thresholds for qualification of an arriving task for low ($\lambda = 10$) and high ($\lambda = 100$) arrival rates of the requests. It can be observed in general that the qualification thresholds decrease as the time approaches towards the terminal time. This is due to the fact that the valuable option of allocating an available VM to a higher complexity task reduces in probability. Furthermore, as we approach the horizon, it is more valuable to allocate a VM to a lower complexity task than to not allocate it at all. It can also be observed from Fig.~\ref{fig:low_rate} that for lower arrival rates, the thresholds drop quickly as compared to the thresholds for the high arrival rates in Fig.~\ref{fig:high_rate}. This is because the expected arrivals are lower in the former and hence the mechanism adjusts the thresholds to qualify more arrivals to tap the revenue potential. The associated pricing curves follow a similar trend as the allocation thresholds except that they are scaled by the characteristics of the VMs. However, in the considered situation, they are identical since $q=1$.

%\begin{figure}
%\centering
%\begin{subfigure}{\linewidth}
%  \centering
%\includegraphics[width=3.2in]{Figures/stoc_trace.eps}
%  \caption{Stochastic arrival of tasks with random computational complexities and optimal allocation thresholds.}
%  \label{fig:stoc_trace}
%\end{subfigure}\\
%\begin{subfigure}{\linewidth}
%  \centering
%\includegraphics[width=3.2in]{Figures/rev.eps}
%\caption{Revenue generated by the CSP over time.}
%  \label{fig:rev}
%\end{subfigure}
%\caption{Application of the proposed algorithm on an example realization.\vspace{-0.2in}}
%\label{fig:experiment_fig}
%\end{figure}

%In order to illustrate the execution of the proposed algorithm, we provide an example realization over a finite horizon of $3$ hours. We assume a task arrival rate of $\lambda = 50$ requests per hour and $N_0 = 50$ VMs. The obtained results are presented in Fig.~\ref{fig:experiment_fig}. Fig.~\ref{fig:stoc_trace} shows a trace of the sequential arrival of computational requests with random complexities. The blue impulses represents the task complexities while the black lines illustrate adaptive and dynamic qualification threshold. Note that each time a task qualifies for allocation, the qualification threshold rises as a result of the revenue maximization mechanism. This behaviour can be directly explained using Theorem~\ref{theorem2}. Fig.~\ref{fig:rev} shows the revenue generated by the CSP corresponding to the optimal VM allocation in the example. The revenue maximizing framework ensures that the expected revenue generated at the terminal time is the highest.

\begin{figure}[t!]
  \centering
  \includegraphics[width=2.7in]{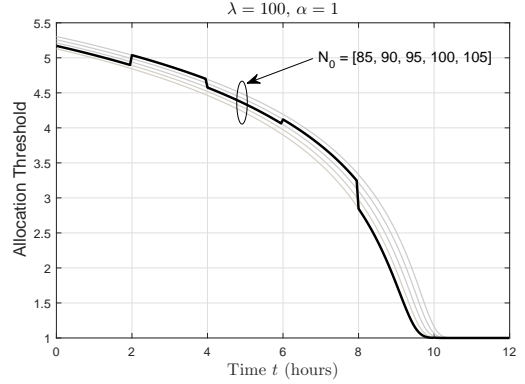}\\
  \caption{Optimal allocation threshold for varying number of available VMs.\vspace{-0.3in}}\label{fig1}
\end{figure}

Next we investigate the adaptive and resilient behaviour of the proposed mechanism.
%We assume that the CSP has 100 VMs available initially, i.e., $N_0 = 100$.
A set of failures and capacity enhancements are simulated at fixed times. For instance, a loss of $15$ VMs and $5$ VMs is assumed to occur at $t = 2$ hours and $t = 6$ hours respectively. Similarly, new additions of $10$ VMs and $5$ VMs is assumed to occur at $t = 4$ hours and $t = 8$ hours. Note that when a loss of $15$ VMs occurred at $t=2$ hours, the situation becomes equivalenlt to as if initially the CSP had $85$ available VMs. Therefore from $t = 2$ onwards, the optimal revenue maximizing policy is to use the threshold and pricing corresponding to the $N_0 = 85$ curve. As the number of available VMs change over times, the optimal policy needs to be updated. The optimal dynamic allocation policy for the above mentioned events is shown by the bold line in Fig.~\ref{fig1}. Notice that as the number of available VMs decreases, the allocation threshold and the price jumps in order to make up for the lost revenue. Similarly, if new VMs become available, the threshold and prices drop in order to strike a new balance between the qualifying tasks and the payment.

\begin{figure}[t!]
  \centering
  \includegraphics[width=2.7in]{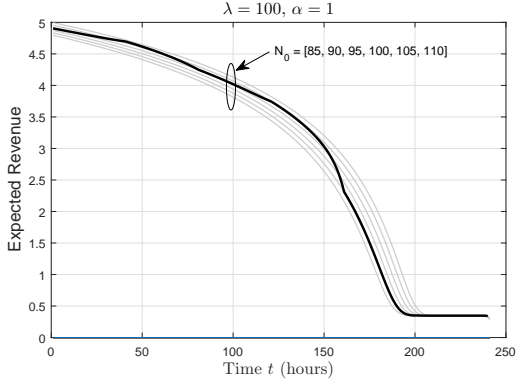}\\
  \caption{Expected Revenue of the CSP over time under variations in the number of available VMs.\vspace{-0.3in}}\label{fig:revenue}
\end{figure}

In Fig.~\ref{fig:revenue}, we show the behaviour of the adaptive and resilient mechanism on the expected revenue of the CSP in response to the variations in the number of available VMs at the cloud server. It can be observed that the adaptive strategy is able to maintain a high expected revenue despite variations in the number of available VMs. Note that during the times when there is a drop in the number of available VMs, the expected revenue does not fall as much due to a rectified allocation and pricing policy as illustrated in Fig.~\ref{fig1}. Hence, it is shown that a timely rectification of allocation and pricing decision enables the mechanism to be adaptive and resilient against any significant changes in the available resources to the CSP.

%\begin{figure}
%  \centering
%  % Requires \usepackage{graphicx}
%  \includegraphics[width=3in]{Figures/comparison.eps}
%  \caption{Comparison of the expected revenue.\vspace{-0.2in}}\label{Fig:comparison}
%\end{figure}

%Finally, in Fig.~\ref{Fig:comparison}, we provide a comparison of the expected revenue achieved using the proposed algorithm with two benchmark strategies, i.e., the random allocation (using a coin toss to allocate a VM to each request) and the first-come-first-served strategy. Although the random strategy appears to achieve higher expected revenue for low density of requests due to the greedy nature of the strategy. However, the maximum achievable revenue saturates rapidly as the density increases or if the allocation period is increased. For higher density of requests, the proposed mechanism achieves a significantly higher expected revenue.

%As an example, the qualification thresholds for the case of low and high arrival rate of tasks are provided in Fig.~\ref{fig:thresholds}. It can be observed that

%\begin{figure}[h!]
%  \centering
%  % Requires \usepackage{graphicx}
%  \includegraphics[width=3.4in]{Figures/stoc_trace.eps}
%  \caption{Trace of experiment.}\label{fig:stoc_trace}
%\end{figure}

\vspace{-0.1cm}
\section{Conclusion}\label{Sec:Conclusion}
In this paper, we have proposed an adaptive and resilient dynamic revenue maximizing framework for cloud computing environments. The framework uses a threshold-based filtering policy making real time allocation and pricing decisions for sequentially arriving computational requests.
%Optimal policies have been derived that balance the tradeoff between the number of qualifying requests and the contribution towards the revenue.
%It has been shown that the optimal allocation thresholds are independent of the characteristics of the available VMs at the cloud server. In fact, they only depend on the number of available VMs, the arrival rate of the requests as well as the average computational complexity of the tasks.
%Furthermore,
It has been shown that the framework is adaptive and resilient to changes in the number of available VMs or the statistical properties of the arrivals. The set of optimal policies can be pre-computed and used as a lookup table as conditions at the cloud server change over time. Therefore, the developed framework provides an optimal and implementable mechanism for allocation and pricing in cloud computing environments. Future directions in this line of work may include developing optimal allocation policies for multiple types of identical resources available at the CSP. Furthermore, the allocation framework can be extended to multiple layers such as in fog/edge computing paradigms.

%%%%%%%%%%%%%%%%%%%%%%%%%%%%%%%%%%%%%%%%%%%%%%%%%%%%%%%%%%%%%%%%%%%%%%%%%%%%%%%%

%%%%%%%%%%%%%%%%%%%%%%%%%%%%%%%%%%%%%%%%%%%%%%%%%%%%%%%%%%%%%%%%%%%%%%%%%%%%%%%%

%%%%%%%%%%%%%%%%%%%%%%%%%%%%%%%%%%%%%%%%%%%%%%%%%%%%%%%%%%%%%%%%%%%%%%%%%%%%%%%%
%\section*{APPENDIX}

\appendices

\section{Proof of Theorem~\ref{main_theorem}} \label{proof_main_theorem}
\begin{figure*}
{\small
\begin{align} \label{L_H}
&\frac{\partial \mathcal{L}(s,H(s), H^{\prime}(s))}{\partial H(s)} = \mathcal{L}_{H} = \sum_{n=1}^{N} \frac{1}{(n-1)!} e^{-H(s)} F_X^{-1}\left(1 - \frac{H^{\prime}(s)}{\lambda}\right)H^{\prime}(s) \left((n-1)H^{n-2}(s) - H^{n-1}(s)\right),
\end{align} \vspace{-0.3cm}
%\begin{align}\label{L_H_prime}
%&\frac{\partial L(s,H(s), H^{\prime}(s))}{\partial H^{\prime}(s)} = \mathcal{L}_{H^{\prime}} =  \sum_{n=1}^{N} \frac{1}{(n-1)!} e^{-H(s)} H^{n-1}(s) \left( F_X^{-1}\left(1 - \frac{H^{\prime}(s)}{\lambda}\right) - \frac{H^{\prime}(s)}{\lambda f_X \left( F_X^{-1} \left(1 -  \frac{H^{\prime}(s)}{\lambda}\right)\right)} \right) ,
%\end{align}
\begin{align}\label{d_dt_H_prime}
\frac{d}{dt}  \mathcal{L}_{H^{\prime}} &= \sum_{n=1}^{N} \frac{1}{(n-1)!} \left[ - 2 \frac{H^{n-1}(s)H^{\prime \prime}(s)}{\lambda f_X \left( F_X^{-1}\left(1 - \frac{H^{\prime}(s)}{\lambda}\right) \right)}  -   H^{\prime}(s)H^{n-1}(s)F_X^{-1}\left(1 - \frac{H^{\prime}(s)}{\lambda}\right) + \right.\notag \\ &\left.
 (n-1)H^{n-1}(s)H^{\prime}(s)F_X^{-1}\left(1 - \frac{H^{\prime}(s)}{\lambda}\right) - \frac{H^{n-1}(s)(H^{\prime}(s))^2}{\lambda F_X^{-1}\left(1 - \frac{H^{\prime}(s)}{\lambda}\right)} -   \frac{H^{n-1}(s) H^{\prime}(s) H^{\prime \prime}(s) f_X^{\prime}(F_X^{-1}(1 - \frac{H^{\prime}(s)}{\lambda}))}{\lambda^2 f^3_X \left( F_X^{-1}\left(1 - \frac{H^{\prime}(s)}{\lambda}\right) \right)}  \right].
\end{align}
}
\end{figure*}
\begin{figure*}
\vspace{-0.6cm}
\begin{align}\label{E_L_equation}
%\mathcal{L}_H - \frac{d}{dt} \mathcal{L}_{H^{\prime}} =
\hspace{-0.1cm} \sum_{n=1}^{N} \frac{1}{(n-1)!} \hspace{-0.1cm} \left[ 2 H^{n-1}(s)H^{\prime \prime}(s)  \hspace{-0.05cm} - \hspace{-0.05cm}  (H^{\prime}(s))^2 \left( H^{n-1}(s) \hspace{-0.05cm} - \hspace{-0.05cm} (n \hspace{-0.05cm} - \hspace{-0.05cm} 1)H^{n-2}(s)\right) + \frac{H^{n-1}(s) H^{\prime}(s) H^{\prime \prime}(s) f_X^{\prime}(F_X^{-1}(1 - \frac{H^{\prime}(s)}{\lambda}))}{\lambda f^2_X \left( F_X^{-1}\left(1 - \frac{H^{\prime}(s)}{\lambda}\right) \right)}  \right] \hspace{-0.1cm} = \hspace{-0.05cm} 0.
\end{align}\vspace{-0.4cm}
\begin{align}\label{exponential_threshold}
y_{N_t}(t) = \frac{1}{\alpha} + \frac{\lambda}{\alpha}\int_t^T  e^{-\alpha y_{N_t}(s)} \frac{ \sum_{n=1}^{N_t} \frac{1}{(n-1)!} \left[ \left(\int_t^s \lambda e^{-\alpha y_{N_t}(s)}du\right)^{n-1}  -(n-1) \left(\int_t^s \lambda e^{-\alpha y_{N_t}(s)}du\right)^{n-2} \right] }{ \sum_{n=1}^{N_t} \frac{1}{(n-1)!} \left(\int_t^s \lambda e^{-\alpha y_{N_t}(s)}du\right)^{n-1} } ds.
\end{align}
\hrule
\vspace{-0.6cm}
\end{figure*}
%\noindent The total expected revenue if $N_t$ VMs are available at time~$t$ can be expressed as follows:\vspace{-0.4cm}
%
%{\small
%\begin{align}
%\mathcal{R}(\{q\}_{N_t},t) &= \hspace{-0.1cm}\sum_{n=1}^{N_t}   q \hspace{-0.1cm} \int_t^T  \hspace{-0.2cm} y_{N_t}(s) \hat{\lambda}(s) \exp \left(- \hspace{-0.1cm} \int_t^s \hspace{-0.1cm} \hat{\lambda}(u) du \right) \notag \times \\& \hspace{0.8in}\frac{\left(\int_t^s \hat{\lambda}(u) du \right)^{n-1}}{(n-1)!} ds + \kappa_T,
%\end{align}
%\begin{align}
%& \hspace{-0.4cm}=  q \int_t^T  \hspace{-0.1cm} y_{N_t}(s) \hat{\lambda}(s)  \exp \left(- \int_t^s \hspace{-0.1cm} \hat{\lambda}(u) du \right) \notag \\ & \hspace{0.5in}\sum_{n=1}^{N_t} \frac{\left(\int_t^s \hat{\lambda}(u) du \right)^{n-1}}{(n-1)!}ds + \kappa_T.
%\end{align}
%}
%For large $N$, the expected revenue can be further expressed as follows:
%\begin{align}
%R(q,t) \overset{N \rightarrow \infty}{=}  q \int_t^T  \hspace{-0.2cm} y_{N_t}(s) \hat{\lambda}(s) \left(1 + \int_t^s \hat{\lambda}(u) du \right) ds.
%\end{align}
Let $H(s)= \int_t^s \hat{\lambda}(u) du = \int_t^s \lambda\left( 1 - F_X(y_{N_t}(u)) \right) du$. Then the expected revenue at time $t$ if $N_t$ VMs are available can be written as follows:\vspace{-0.4cm}

{\small
\begin{align} \label{expected_revenue_new}
\mathcal{R}(\{q\}_{N_t},t) = &q \hspace{-0.1cm}\int_t^T \hspace{-0.1cm} \sum_{n=1}^{N_t} \frac{1}{(n-1)!} F_X^{-1} \left(  1 - \frac{H^{\prime}(s)}{\lambda}  \right) \times \notag \\&  \qquad H^{\prime}(s) e^{-H(s)} (H(s))^{n-1} ds + \kappa_T.
\end{align}
}
This functional can be optimized for the time varying threshold $y_{N_t}(t)$ using the calculus of variations~\cite{calculus_variations}. We denote the kernel of integration as \vspace{-0.4cm}

{\small
\begin{align}
\mathcal{L}(s,H(s), H^{\prime}(s)) \hspace{-0.05cm} =& \hspace{-0.05cm} \sum_{n=1}^{N_t} \hspace{-0.05cm} \frac{H^{\prime} (s) e^{-H(s)} H^{n-1}(s)}{(n-1)!} F_X^{-1} \hspace{-0.1cm} \left(  \hspace{-0.05cm} 1 \hspace{-0.05cm}- \frac{H^{\prime}(s)}{\lambda} \hspace{-0.05cm}  \right).
\end{align}
}
The Euler-Lagrange equation~\cite{calculus_variations} represents the necessary condition satisfied by $H(s)$ to be a stationary function of the expected revenue $R(\{q\}_{N_t},t)$ and can be written as follows: \vspace{-0.4cm}

{\small
\begin{align} \label{Euler_lagrange}
\frac{\partial \mathcal{L}(s,H(s), H^{\prime}(s))}{\partial H(s)} - \frac{d}{dt} \frac{\partial \mathcal{L}(s,H(s), H^{\prime}(s))}{\partial H^{\prime}(s)} = 0.
\end{align}
}
The partial derivatives and the condition satisfied by the resulting Euler-Lagrange equation are given by \eqref{L_H}, \eqref{d_dt_H_prime}, and \eqref{E_L_equation} respectively. The expression in~\eqref{E_L_equation} can be further reduced as follows: \vspace{-0.4cm}

{\small
\begin{align}
&  2 H^{\prime \prime}(s) \hspace{-0.05cm} - \hspace{-0.05cm} (H^{\prime}(s))^2 \frac{\sum_{n=1}^{N_t} \hspace{-0.1cm}\frac{1}{(n-1)!} \hspace{-0.1cm}\left( H^{n-1}(s) - (n-1)H^{n-2}(s)\right)}{\sum_{n=1}^{N_t} \frac{1}{(n-1)!} H^{n-1}(s) }  +  \notag \\ &\frac{ H^{\prime}(s) H^{\prime \prime}(s) f_X^{\prime}(F_X^{-1}(1 - \frac{H^{\prime}(s)}{\lambda}))}{\lambda f^2_X \left( F_X^{-1}\left(1 - \frac{H^{\prime}(s)}{\lambda}\right) \right)}  = 0.
\end{align}
}
Let $J_{N_t}(t,s) = \frac{\sum_{n=1}^{N_t} \frac{1}{(n-1)!} \left( H^{n-1}(s) - (n-1)H^{n-2}(s)\right)}{\sum_{n=1}^{N_t} \frac{1}{(n-1)!}  H^{n-1}(s) }$. Then, plugging back $H(s) = \int_t^s \lambda \left( 1 - F_X(y_{N_t}(u)) \right) du$ results in the following: \vspace{-0.5cm}

{\small
\begin{align}
%&-2 f(y_{N_t}(s))  y_{N_t}^{\prime}(s) - \lambda (1 - F_X(y_{N_t}(s)))^2 J_{N_t}(t,s) - \notag \\ &\frac{(1 - F_X(y_{N_t}(s))) y_{N_t}^{\prime}(s) f^{\prime}_X(y_{N_t}(s))}{f_X(y_{N_t}(s))} = 0, \notag \\
&-2  y_{N_t}^{\prime}(s) - \frac{\lambda (1 - F_X(y_{N_t}(s)))^2 J_{N_t}(t,s)}{f(y_{N_t}(s))} - \notag \\ &\frac{(1 - F_X(y_{N_t}(s))) y_{N_t}^{\prime}(s) f^{\prime}_X(y_{N_t}(s))}{f^2_X(y_{N_t}(s))} = 0
\end{align}
}
It can be further expressed as follows: \vspace{-0.5cm}

{\small
\begin{align}\label{cond}
&- y_{N_t}^{\prime}(s) - y_{N_t}^{\prime}(s) \left( 1 + \frac{(1 - F_X(y_{N_t}(s))) f^{\prime}_X(s)}{(f_X(y_{N_t}(s)))^2} \right) = \notag \\ &  \frac{\lambda (1 - F_X(y_{N_t}(s)))^2 J_{N_t}(t,s)}{f_X(y_{N_t}(s))},
\end{align}
}
Since \\$\frac{d}{ds} \left( \frac{1 - F_X(y_{N_t}(s)}{f(y_{N_t}(s))} \right) \hspace{-0.1cm}= -y^{\prime}_{1}(s) \left( 1 + \frac{(1 - F_X(y_{N_t}(s))) f_X^{\prime}(y_{N_t}(s))}{(f_X(y_{N_t}(s)))^2} \right)$, so the condition in~\eqref{cond} can be written as follows: \vspace{-0.4cm}

{\small
\begin{align}
\frac{d}{ds} \hspace{-0.05cm} \left( \hspace{-0.05cm} \frac{1 - F_X(y_{N_t}(s))}{f_X(y_{N_t}(s))} \hspace{-0.05cm} \right) \hspace{-0.05cm} =& y^{\prime}_{1}(s) + \lambda \frac{(1 - F_X(y_{N_t}(s)))^2 }{f_X(y_{N_t}(s))}J_{N_t}(t,s) .
\end{align}
}
Integrating both sides with respect to $s$ from $t$ to $T$ results in the following: \vspace{-0.4cm}

{\small
\begin{align}
&\left( \frac{1 - F_X(y_{N_t}(T))}{f_X(y_{N_t}(T))} \right) - \left( \frac{1 - F_X(y_{N_t}(t))}{f_X(y_{N_t}(t))} \right) = y_{N_t}(T) - \notag \\ & y_{N_t}(t) +   \lambda \int_t^T  \frac{(1 - F_X(y_{N_t}(s)))^2 }{f_X(y_{N_t}(s))}J_{N_t}(t,s) ds.
\end{align}
}
Using the boundary condition, $y_{N_t}(T) = \frac{1 - F_X(y_{N_t}(T))}{f(y_{N_t})(T)}$, i.e., at the terminal time only the virtual valuation of the users can be recovered, it follows that the cutoff curve $y_{N_t}(t)$ satisfies the equation given by Theorem~\ref{main_theorem}.
%following condition: \vspace{-0.4cm}
%
%{\small
%\begin{align}
%y_{N_t}(t) = \frac{1 - F_X(y_{N_t}(t))}{f_X(y_{N_t}(t))} +  &\lambda \int_t^T \hspace{-0.1cm}\frac{(1 - F_X(y_{N_t}(s)))^2 }{f_X(y_{N_t}(s))} J_{N_t}(t,s)ds,
%\end{align}
%}
%which completes the proof.

\vspace{-0.1cm}
\section{Proof of Theorem~\ref{theorem2}\vspace{-0.1cm}}~\label{proof_theorem2}
First we need to show that for $\{M_t,N_t \in \mathbb{Z}^+ : M_t \geq N_t\}$, $J_{M_t}(t,s) \leq J_{M_t}(t,s), \forall t,s$. To do this we will show that $J_{N_{t} + 1}(t,s) \leq J_{N_t}(t,s)$. It is equivalent to showing that $J_{N_{t} + 1}(t,s) - J_{N_t}(t,s)$$\leq 0$, i.e., \vspace{-0.41cm}

{\small
\begin{align}
\frac{   \sum_{n=1}^{N_t + 1} \frac{H^{n-1}(s)}{(n-1)!}   -  \sum_{n=1}^{N_t+1} \frac{(n-1)H^{n-2}(s)}{(n-1)!}}{\sum_{n=1}^{N_t+1} \frac{H^{n-1}(s)}{(n-1)!}}   -  \notag \\ \frac{   \sum_{n=1}^{N_t} \frac{H^{n-1}(s)}{(n-1)!}   -  \sum_{n=1}^{N_t} \frac{(n-1)H^{n-2}(s)}{(n-1)!}}{\sum_{n=1}^{N_t} \frac{H^{n-1}(s)}{(n-1)!}} \leq 0.
\end{align}
}
It can be further expressed as follows: \vspace{-0.41cm}

{\small
\begin{align}
\left( \sum_{n=1}^{N_t} \frac{H^{n-1}(s)}{(n-1)!} \right) \left(\frac{H^{N_t}(s) - N_t H^{N_t-1}(s)}{N_t!}\right) - \notag\\ \left( \sum_{n=1}^{N_t} \frac{H^{n-1}(s) - (n-1)H^{n-2}(s)}{(n-1)!}\right)\frac{H^{N_t}(s)}{N_t!} \leq 0.
\end{align}
}
Expanding the condition results in the following: \vspace{-0.5cm}

{\small
\begin{align}
&\sum_{n=1}^{N_t} \left( \frac{H^{N_t + n -1}(s) - N_t H^{N_t + n -2}(s) - H^{N_t + n - 1}(s) }{(n-1)!} + \right. \notag \\& \hspace{1.3in} \left. \frac{(n-1)H^{N_t + n -2}(s)}{(n-1)!} \right)\leq 0.
\end{align}
}
It is equivalent to the following condition:\vspace{-0.5cm}

{\small
\begin{align}
\sum_{n=1}^{N_t} \frac{H^{N_t + n -2}(s)(n+1-N_t)}{(n-1)} \leq 0,
\end{align}
}
which is true since $(n+1-N_t) \leq 0, \forall n = 1, \ldots, N_t$. Therefore, it is evident that $J_{N_{t} + 1}(t,s) \leq J_{N_t}(t,s)$. Using induction it can be shown that the inequality $J_{M_t}(t,s) \leq J_{N_t}(t,s)$ holds for general $M_t$ and $N_t$ such that $M_t \geq N_t, \forall t$. From Theorem~\ref{main_theorem}, the result follows directly with the assumption of increasing virtual valuations, i.e., $x - \frac{1 - F_X(x)}{f_X(x)}$ is increasing in $x$.

%\section*{ACKNOWLEDGMENT}
%
%The preferred spelling of the word ÒacknowledgmentÓ in America is without an ÒeÓ after the ÒgÓ. Avoid the stilted expression, ÒOne of us (R. B. G.) thanks . . .Ó  Instead, try ÒR. B. G. thanksÓ. Put sponsor acknowledgments in the unnumbered footnote on the first page.

%%%%%%%%%%%%%%%%%%%%%%%%%%%%%%%%%%%%%%%%%%%%%%%%%%%%%%%%%%%%%%%%%%%%%%%%%%%%%%%%

\bibliographystyle{IEEEtran}
\bibliography{references}

\begin{thebibliography}{10}
\providecommand{\url}[1]{#1}
\csname url@samestyle\endcsname
\providecommand{\newblock}{\relax}
\providecommand{\bibinfo}[2]{#2}
\providecommand{\BIBentrySTDinterwordspacing}{\spaceskip=0pt\relax}
\providecommand{\BIBentryALTinterwordstretchfactor}{4}
\providecommand{\BIBentryALTinterwordspacing}{\spaceskip=\fontdimen2\font plus
\BIBentryALTinterwordstretchfactor\fontdimen3\font minus
  \fontdimen4\font\relax}
\providecommand{\BIBforeignlanguage}[2]{{%
\expandafter\ifx\csname l@#1\endcsname\relax
\typeout{** WARNING: IEEEtran.bst: No hyphenation pattern has been}%
\typeout{** loaded for the language `#1'. Using the pattern for}%
\typeout{** the default language instead.}%
\else
\language=\csname l@#1\endcsname
\fi
#2}}
\providecommand{\BIBdecl}{\relax}
\BIBdecl

\bibitem{zhu}
Q.~Zhang, Q.~Zhu, M.~F. Zhani, R.~Boutaba, and J.~L. Hellerstein, ``Dynamic
  service placement in geographically distributed clouds,'' \emph{IEEE Journal
  on Selected Areas in Communications}, vol.~31, no.~12, pp. 762--772, 2013.

\bibitem{iot}
A.~Al-Fuqaha, M.~Guizani, M.~Mohammadi, M.~Aledhari, and M.~Ayyash, ``Internet
  of things: A survey on enabling technologies, protocols, and applications,''
  \emph{IEEE Commun. Surveys Tuts.}, vol.~17, no.~4, pp. 2347--2376,
  Fourthquarter 2015.

\bibitem{twc}
M.~J. Farooq and Q.~Zhu, ``On the secure and reconfigurable multi-layer network
  design for critical information dissemination in the internet of battlefield
  things {(IoBT)},'' \emph{IEEE Trans. Wireless Commun.}, vol.~PP, no.~99, pp.
  1--1, 2018.

\bibitem{massive_iot}
M.~J. Farooq, H.~ElSawy, Q.~Zhu, and M.~S. Alouini, ``Optimizing mission
  critical data dissemination in massive {IoT} networks,'' in \emph{15th Int.
  Symp. Model. Opt. Mobile, Ad Hoc, and Wireless Netw. (WiOpt 2017)}, May 2017,
  pp. 1--6.

\bibitem{resource_allocation_cloud}
W.~Song, Z.~Xiao, and Q.~Chen, ``Dynamic resource allocation using virtual
  machines for cloud computing environment,'' \emph{IEEE Trans. Parallel
  Distrib. Syst.}, vol.~24, pp. 1107--1117, 2013.

\bibitem{cloud_performance}
S.~Marrone and R.~Nardone, ``Automatic resource allocation for high
  availability cloud services,'' \emph{Procedia Computer Science}, vol.~52, pp.
  980 -- 987, 2015.

\bibitem{risk_based}
Y.~Lu, S.~T. Maguluri, M.~S. Squillante, and C.~W. Wu, ``Risk-based dynamic
  allocation of computing resources,'' \emph{SIGMETRICS Perform. Eval. Rev.},
  vol.~44, no.~2, pp. 27--29, Sep. 2016.

\bibitem{cloud_pricing}
H.~Xu and B.~Li, ``Dynamic cloud pricing for revenue maximization,'' \emph{IEEE
  Trans. Cloud Comput.}, vol.~1, no.~2, pp. 158--171, July 2013.

\bibitem{spot_markets}
Q.~Zhang, Q.~Zhu, and R.~Boutaba, ``Dynamic resource allocation for spot
  markets in cloud computing environments,'' in \emph{4th IEEE Int. Conf. Util.
  Cloud Comput.}, Dec 2011, pp. 178--185.

\bibitem{cloud_auction}
H.~Zhang, B.~Li, H.~Jiang, F.~Liu, A.~V. Vasilakos, and J.~Liu, ``A framework
  for truthful online auctions in cloud computing with heterogeneous user
  demands,'' in \emph{2013 Proceedings IEEE INFOCOM}, April 2013, pp.
  1510--1518.

\bibitem{albright}
S.~C. Albright, ``Optimal sequential assignments with random arrival times,''
  \emph{Management Science}, vol.~21, no.~1, pp. 60--67, 1974.

\bibitem{revenue_maximization}
A.~Gershkov and B.~Moldovanu, ``Dynamic revenue maximization with heterogeneous
  objects: A mechanism design approach,'' \emph{American Economic Journal:
  Microeconomics}, vol.~1, no.~2, pp. 168--198, 2009.

\bibitem{hard_book}
G.~H. Hardy, J.~E. Littlewood, and G.~Polya, \emph{Inequalities}.\hskip 1em
  plus 0.5em minus 0.4em\relax Cambridge UK: Cambridge University Press, 1934.

\bibitem{stochastic_book}
S.~M. Ross, \emph{Stochastic Processes}, 1996.

\bibitem{Myerson:1981:OAD:2781650.2781656}
R.~B. Myerson, ``Optimal auction design,'' \emph{Math. Oper. Res.}, vol.~6,
  no.~1, pp. 58--73, Feb. 1981.

\bibitem{cloud_attack}
N.~V. Juliadotter and K.~K.~R. Choo, ``Cloud attack and risk assessment
  taxonomy,'' \emph{IEEE Cloud Comput.}, vol.~2, no.~1, pp. 14--20, Jan 2015.

\bibitem{fixed_point}
R.~L. Burden, J.~D. Faires, and A.~M. Burden, \emph{Numerical Analysis}, 1985.

\bibitem{calculus_variations}
I.~M. Gelfand and S.~V. Fomin, \emph{{Calculus of Variations (Dover Books on
  Mathematics)}}.\hskip 1em plus 0.5em minus 0.4em\relax Dover Publications,
  Oct. 2000.

\end{thebibliography}

\end{document}